\documentclass[prd,reprint,nofootinbib,superscriptaddress,preprintnumbers,showpacs]{revtex4-1}
\usepackage{amsfonts}
\usepackage{amssymb}
\usepackage{amsmath}
\usepackage{graphicx,color}
\usepackage{dcolumn}
\usepackage{epsfig}
\usepackage{bm}
\usepackage{bbm}
\usepackage{ulem}
\usepackage{hyperref}
\usepackage{lineno}

\usepackage{color}

\def\gtap{\ \raise.3ex\hbox{$>$\kern-.75em\lower1ex\hbox{$\sim$}}\ }
\def\ltap{\ \raise.3ex\hbox{$<$\kern-.75em\lower1ex\hbox{$\sim$}}\ }
\begin{document}

\title{
Triangle singularity
appearing as an $X(3872)$-like peak
in $B\to (J/\psi\pi^+\pi^-) K\pi$
}
\author{Satoshi X. Nakamura}
\email{satoshi@ustc.edu.cn}
\affiliation{
University of Science and Technology of China, Hefei 230026, 
People's Republic of China
}
\affiliation{
State Key Laboratory of Particle Detection and Electronics (IHEP-USTC), Hefei 230036, People's Republic of China}

\begin{abstract}
We consider a triangle diagram for 
$B^0\to (J/\psi\pi^+\pi^-) K^+\pi^-$ where an $X(3872)$ peak
has been observed experimentally.
 We demonstrate that a triangle singularity
 inherent in the triangle
 diagram creates a sharp peak in the $J/\psi\pi^+\pi^-$ invariant mass
 distribution
when the  final $(J/\psi\pi^+\pi^-)\pi$
invariant mass is at and around the $D^*\bar D^*$ threshold.
 The position and width of the peak is
 3871.68~MeV (a few keV above the $D^{*0}\bar{D}^0$ threshold)
 and $\sim$0.4~MeV, respectively, in perfect agreement with
 the precisely measured $X(3872)$ mass and width:
$3871.69\pm 0.17$~MeV and $< 1.2$~MeV.
 This remarkable agreement is virtually parameter-free.
The result indicates that 
the considered mechanism has to be understood in advance
when separating an $X(3872)$-pole contribution from
$B^0\to (J/\psi\pi^+\pi^-) K^+\pi^-$ data;
the separation yields 
an $X(3872)\pi$ lineshape that could be used to determine the
$X(3872)$ mass.
We suggest
a method to set a constraint on the triangle mechanism by analyzing a charge
 analogous process $B^0\to (J/\psi\pi^0\pi^-) K^+\pi^0$ where a similar
 triangle singularity generates an $X^-(3876)$-like peak.
\end{abstract}

\maketitle

\section{Introduction}
\label{sec:intro}
Establishing the existence of exotic hadrons, which are not accommodated
by the conventional quark model picture~\cite{qm},
is arguably the most prioritized problem in the contemporary
hadron spectroscopy.
The discovery of $X(3872)$~\cite{belle_x3872_jpsi-rho}
triggered this trend where
the nature of $X(3872)$ has always been the central problem; see
Refs.~\cite{review_swanson,review_voloshin,review_chen,review_hosaka,review_lebed,review_esposito,review_ali,review_guo,review_olsen,review_raphael}
for reviews.
Experimentally, $X(3872)$ has been observed
not only in $B$ meson decays where it was discovered~\cite{belle_x3872_jpsi-rho,babar_x3872_jpsi-rho},
but also in $pp$ and $\bar{p}p$ collisions~\cite{cdf_x3872_jpsi-rho,d0_x3872_jpsi-rho,lhcb_x3872_jpsi-rho}
 and $e^+e^-$ annihilations~\cite{bes3_x3872_jpsi-rho}.
$X(3872)$ has been confirmed to decay into several channels such as
$J/\psi\rho^0(\rho^0\to\pi^+\pi^-)$~\cite{belle_x3872_jpsi-rho,babar_x3872_jpsi-rho,cdf_x3872_jpsi-rho,d0_x3872_jpsi-rho,lhcb_x3872_jpsi-rho,bes3_x3872_jpsi-rho},
$J/\psi\omega$~\cite{babar_x3872_jpsi-omega,bes3_x3872_jpsi-omega},
$J/\psi\gamma$~\cite{belle_x3872_jpsi-gamma},
$D^{*0}\bar{D}^0$~\cite{belle_x3872_ddstar}, and more.
\begin{figure*}[t]
\begin{center}
\includegraphics[width=1\textwidth]{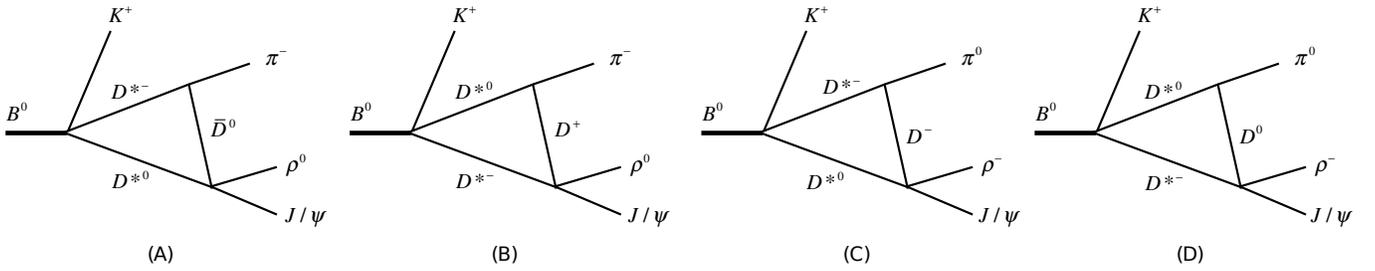}
\end{center}
 \caption{Triangle diagrams (A,B) [(C,D)]
 contributing to $B^{0}\to J/\psi \rho^0 K^+\pi^-$
 [$B^{0}\to J/\psi \rho^- K^+\pi^0$].
 The diagrams (A,C,D) cause triangle singularities
to generate a sharp peak in
 $J/\psi \rho$ invariant mass ($M_{J/\psi \rho}$) distributions
 at $M_{J/\psi \rho}\sim 3.872, 3.876$ and 3.875~GeV, respectively.
From the left to right, we refer to the diagrams in the text as diagrams~A, B, C, and D, respectively.
 }
\label{fig:diag}
\end{figure*}

Many theoretical attempts have been made to understand 
what $X(3872)$ consists of.
Because of the close proximity of its mass to the
$D^{*0}\bar{D}^0$~\footnote{Charge conjugates are implied throughout.} threshold,
a $D^{*0}\bar{D}^0$ molecule is a popular idea~\cite{swanson_molecule,zhao_molecule}.
However, a pure molecule picture makes it difficult to understand its
formation rate in the hadron collider experiments~\cite{suzuki}.
Thus a superposition of the molecule with an excited charmonium is
considered more plausible~\cite{suzuki,Kalashnikova,takizawa}.
The latest Lattice QCD~\cite{lqcd_Prelovsek,lqcd_Padmanath}
found a state that could be identified with 
$X(3872)$, and disfavored diquark-antidiquark 
interpretations~\cite{tetra_maiani,tetra_chen}.
Yet, it seems difficult to reach a 
consensus on the structure of $X(3872)$ within the so-far proposed ideas
(maybe except for Lattice QCD) because of
lots of unknowns concerning the relevant hadron dynamics,
and one may fine-tune them to reproduce available data.

Another issue is that 
a spectrum peak of $X(3872)$ could be partly faked by a kinematical
effect, the triangle singularity (TS) in particular.
A TS occurs from a triangle diagram like
Fig.~\ref{fig:diag} only if a special kinematical condition is realized:
all three internal particles are simultaneously on-shell and their
momenta are collinear like in a classical process~\cite{landau,coleman,s-matrix}. 
The TS can significantly enhance the amplitude,
and can show up as a bump in, for example, $J/\psi \rho$ and $J/\psi \rho\pi$
invariant mass distributions of the processes in Fig.~\ref{fig:diag}.
For mathematical details and practical use, we refer the readers to Ref.~\cite{TS-Pc2}.
Attempts have been made to interpret 
some $XYZ$ exotic candidates as bumps
due to TSs~\cite{ts1_z3900,ts2_z3900,szczepaniak,xhliu1,xhliu2,xhliu3,ts4_z3900,ts3_z3900,ts_zc4430,ts_z4050,ts_review}.

The triangle diagram of Fig.~\ref{fig:diag}(A)~\footnote{
Hereafter, we refer to the triangle diagrams of Figs.~\ref{fig:diag}(A),
\ref{fig:diag}(B),
\ref{fig:diag}(C), and \ref{fig:diag}(D) as 
diagrams~A, B, C, and D, respectively.}
 for $B^0\to J/\psi \rho^0 K^+ \pi^-$
satisfies conditions to create an $X(3872)$-like peak
in the $J/\psi \rho^0$ invariant mass distribution
when the $J/\psi \rho^0\pi^-$ invariant mass is at and around the $D^{*-}D^{*0}$
threshold.
Experimentally, an $X(3872)$ peak has been observed in 
the $J/\psi\pi^+\pi^-$ lineshape of
$B^0\to (J/\psi\pi^+\pi^-) K^+ \pi^-$~\cite{belle_x3872kpi,belle_x3872kpi2}~\footnote{
An $X(3872)$ peak
has been also observed
in $B^+\to J/\psi \rho^0 K^0 \pi^+$~\cite{belle_x3872kpi}
for which a similar triangle diagram generates
an $X(3872)$-like peak.
Because of the similarity, we study only the $B^0$ decay processes in 
Fig.~\ref{fig:diag} in this work.
}.
In this work, 
we demonstrate that the TS inherent in the triangle
diagram generates an exactly $X(3872)$-like peak in 
the $J/\psi\pi^+\pi^-$ invariant mass spectrum.
The spectrum peak position and shape agrees with 
the $X(3872)$ mass measured at $\sim$0.01\% precision and
the tightly constrained width
without any fine-tuning of the model parameters. 

Our analysis will indicate that 
the TS should be taken into account 
when studying $X(3872)$
in $B^0\to J/\psi \rho^0 K^+ \pi^-$ in the TS region.
This would be particularly relevant to an idea recently proposed by
Guo~\cite{guo_x3872,sakai_x3872,sakai2_x3872} on
determining the $X(3872)$ mass from
$X(3872)\pi$ and $X(3872)\gamma$ lineshapes.
Characteristic lineshapes are created by TSs
of triangle diagrams similar to
the diagram~A
but different in including an $X(3872)$-pole propagation as
$D^{*0}\bar{D}^0\to X(3872)\to J/\psi\rho^0$~\footnote{
Braaten et al.~\cite{ohio1,ohio2,ohio3} 
also studied similar triangle diagrams
as an amplifier of $X(3872)$ productions.
}.
The idea is based on the fact that 
the lineshapes sensitively change depending on the $X(3872)$ mass.
Because 
$X(3872)\to J/\psi\pi^+\pi^-$ is suppressed by an isospin violation,
the isospin conserving non-resonant process like the diagram~A
could give a comparable $X(3872)$-like contribution as a background for
the $X(3872)$ mass analysis.
In order to 
understand the non-resonant mechanism and 
separate it from the $X(3872)$-pole contribution,
 we will suggest to analyze a charge analogous 
$B^0\to (J/\psi\pi^0\pi^-) K^+ \pi^0$ decay for which TSs from related triangle
diagrams C and D generates an $X^-(3876)$-like peak.

\section{model}
The $B^0\to J/\psi\rho^0 K^+ \pi^-$ amplitude
from the triangle diagram~A
can be written,
in the $J/\psi\rho^0 \pi^-$ center-of-mass (CM) frame,
as
\begin{eqnarray}
 T
  &=& \int d\bm{q}\,
  v_{J/\psi\rho^0;D^{*0}\bar{D}^0}\,
  { 1
  \over
  W - E_{D^{*0}} - E_{\bar{D}^{0}} - E_{\pi^-}
  }
  \nonumber \\
 &&\times  
\Gamma_{\bar{D}^0\pi^-,D^{*-}}
{ 1
\over
  W - E_{D^{*-}} - E_{D^{*0}} }
 V_{K^+D^{*-} D^{*0},B^0}
  \, ,\nonumber\\
  \label{eq:amp}
\end{eqnarray}
where $\bm{q}$ is a loop momentum.
The invariant mass of the $J/\psi\rho^0\pi^-$ subsystem is denoted by $W$,
while the energy of a particle $x$ is 
$E_x$ which depends on the particle mass ($m_x$) and momentum ($\bm{p}_x$) as 
$E_x=\sqrt{\bm{p}^2_x+m^2_x} - i\Gamma_x/2$;
$\Gamma_x$ is the decay width which is nonzero 
for $D^*$.
The summation of intermediate spin states is implied.
We use values from the Particle Data Group (PDG)~\cite{pdg} for the particle masses ($m_x$),
except for the final $\rho$ meson for which our treatment will be
discussed later.
Amplitudes for triangle diagrams B, C, and D
are similar.
In calculating observables for
the $B^0\to J/\psi\rho^0 K^+ \pi^-$ ($B^0\to J/\psi \rho^- K^+ \pi^0$) decay,
the triangle mechanisms A and B (C and D) must be coherently added.
Because of the charge-parity invariance and isospin symmetry of the strong interaction, 
the triangle mechanisms A and B (C and D)
exactly cancel with each other 
in a hypothetical situation where
the charged and neutral $D^{*}$ mesons have the same mass and width.
In reality, the cancellation is incomplete;
the TS peaks are mostly intact while contributions away from the TS region
are largely cancelled.

We emphasize that mass differences between the isospin partners
such as
$({\pi^\pm},{\pi^0})$, $({D^+},{D^0})$, and $({D^{*+}},{D^{*0}})$,
must be taken into account because they
are essentially important whether or not a TS exists in the triangle
diagrams.
Indeed, while the triangle diagrams~A, C, and D
cause TSs, the diagram~B does not.
This is because 
$D^{*0}\to D^+\pi^-$ at on-shell is kinematically forbidden
and thus the kinematical condition for TS is not 
satisfied.
The triangle amplitude of Eq.~(\ref{eq:amp}) for the diagram~A
in the zero-width limit
causes a TS in the kinematical range of:
\begin{eqnarray}
0  &<& M_{J/\psi\rho} - (m_{D^{*0}}+m_{D^{0}}) \leq 0.2~{\rm MeV} \ ,\\
0 &<& W - (m_{D^{*-}}+m_{D^{*0}}) \ltap 1.0~{\rm MeV} \ ,
\label{eq:W-range}
\end{eqnarray}
where $M_{J/\psi\rho}$ denotes the $J/\psi\rho$ invariant mass.
Although finite widths would relax the singularity,
the $D^{*-}$ and $D^{*0}$ widths are very small
as discussed in the next paragraph.
Therefore we expect from the TS a very sharp peak at 
$M_{J/\psi\pi^+\pi^-}\sim m_{D^{*0}}+m_{D^{0}}=3871.7$~MeV that coincides with the $X(3872)$ mass.
Similarly, the TS condition is satisfied 
for the triangle diagram C in 
\begin{eqnarray}
0  &<& M_{J/\psi \rho} - (m_{D^{*0}}+m_{D^{-}}) \leq 0.2~{\rm MeV} \ ,
\label{eq:ts2}
\end{eqnarray}
with $m_{D^{*0}}+m_{D^{-}}=3876.4$~MeV,
and for the diagram D
\begin{eqnarray}
0  &<& M_{J/\psi \rho} - (m_{D^{*-}}+m_{D^{0}}) \leq 0.2~{\rm MeV} \ ,
\label{eq:ts3}
\end{eqnarray}
with $m_{D^{*-}}+m_{D^{0}}=3875.1$~MeV;
the $W$ range is the same as Eq.~(\ref{eq:W-range}).
Thus the coherent sum of the diagrams C and D
is expected to give a sharp $X^-(3876)$-like peak in the
$M_{J/\psi\pi^0\pi^-}$ distribution.

Regarding the $D^{*\pm}$ decay width, we use the central value of the
PDG average, $\Gamma_{D^{*\pm}}=83.4\pm 1.8$~keV~\cite{pdg}.
On the other hand,
the $D^{*0}$ decay width has been given an upper limit only, 
$\Gamma_{D^{*0}}<2.1$~MeV~\cite{pdg}.
Thus we use $\Gamma_{D^{*0}}$ calculated by 
assuming the isospin symmetry between 
$D^{*+} \to D^+\pi^0$ and $D^{*0} \to D^0\pi^0$,
and also by taking account of the experimentally determined branching to
 $D^{*0} \to D^0\gamma$~\cite{pdg}.
We obtain $\Gamma_{D^{*0}}=55$~keV which is very similar to those
derived previously~\cite{guo_x3872,rosner_dstar}.
We use the constant $D^*$ width values in Eq.~(\ref{eq:amp}),
which has been checked to be a very good approximation.

An $s$-wave $D^{*0}\bar{D}^0\to J/\psi\rho^0$ interaction
we use in Eq.~(\ref{eq:amp}) is
\begin{eqnarray}
  v_{J/\psi \rho^0;D^{*0}\bar{D}^0}\,
  = f^{01}_{J/\psi\rho^0}(p) f^{01}_{D^{*0}\bar{D}^0}(p')\,
  \bm{\epsilon}^*_{J/\psi}\times\bm{\epsilon}^*_{\rho}
  \cdot\bm{\epsilon}_{D^{*0}} \ ,
\label{eq:contact}
\end{eqnarray}
where $\bm{\epsilon}_x$ denotes
the polarization vector for a vector meson $x$.
The $J/\psi\rho^0$ pair coming out of this interaction has the spin-parity
$J^P=1^+$ because of the spin combination specified by the interaction.
Thus, if the $J/\psi\rho^0$ pair generates a bump in the invariant mass
($M_{J/\psi\rho}$) distribution, the pair seems like a decay product of a
resonance of $J^P=1^+$, the spin-parity of $X(3872)$.
We have used in Eq.~(\ref{eq:contact})
dipole form factors $f^{LS}_{ij}(p)$ defined by
\begin{eqnarray}
 f^{LS}_{ij}(p) =
g^{LS}_{ij} {p^L\over \sqrt{E_i(p) E_j(p)}}
\left(\frac{\Lambda^2}{\Lambda^2+p^2}\right)^{2+(L/2)}\ ,
\label{eq:ff}
\end{eqnarray}
where $L$ ($S$) is the orbital angular momentum
(total spin) of the $ij$ pair;
$p=|\bm{p}_{ij}|$ with
$\bm{p}_{ij}$ the $i$'s momentum in the $ij$ CM frame.
We use a cutoff $\Lambda=1$~GeV in the form factors
throughout unless otherwise stated.
The coupling strength
($g^{LS}_{ij}$)
for the interaction of Eq.~(\ref{eq:contact}) is little known and thus
left arbitrary.
Microscopically, this contact interaction can be viewed as an axial vector
$D_1$-meson exchange or a quark exchange mechanism~\cite{swanson_molecule}.
An $X(3872)$-pole contribution is not included in $v_{J/\psi\rho^0;D^{*0}\bar{D}^0}$.

The vertex function for $D^{*-}\to \bar{D}^0\pi^-$ is denoted by $\Gamma_{\bar{D}^0\pi^-,D^{*-}}$
in Eq.~(\ref{eq:amp}), and its explicit form is
given in a general form as
\begin{eqnarray}
  \Gamma_{ij,R}(\bm{p}_i,\bm{p}_j;\bm{p}_R)
   &=& \sum_{LS}f^{LS}_{ij}(p_{ij}) (s_is_i^zs_js_j^z|SS^z)
   \nonumber \\
&\times&   (LM S S^z|S_RS^z_R)
   Y_{LM} (\hat{p}_{ij}) \ ,
  \label{eq:vertex}
\end{eqnarray}
where $Y_{LM}$ is spherical harmonics. 
We use a notation of $(abcd|ef)$
as Clebsch-Gordan coefficients in which we write
the spin of a particle $x$ 
$s_x$ and its $z$-component $s_x^z$.
The coupling strength $g^{10}_{\bar{D}^0\pi^-}$ included in the form
factor $f^{10}_{\bar{D}^0\pi^-}$ is determined by fitting the 
partial decay width for $D^{*-}\to \bar{D}^0\pi^-$~\cite{pdg}.

The $B^0\to D^{*-} D^{*0} K^+$ decay vertex
in Eq.~(\ref{eq:amp}) is expressed with two vertex functions
of Eq.~(\ref{eq:vertex}) as
\begin{eqnarray}
  V_{K^+ D^{*-} D^{*0},B^0} &=& 
\exp\left(-b\ {W - m_{D^{*-}}-m_{D^{*0}} \over  m_{D^{*-}}+m_{D^{*0}}}\right)
   \nonumber\\
 &&\times 
\sum_{\cal R}
   \Gamma_{D^{*-} D^{*0},{\cal R}}(\bm{p}_{D^{*-}},\bm{p}_{D^{*0}};\bm{p}_{\cal R})
   \nonumber\\
 &&\times \Gamma_{K^+ {\cal R},B^0}(\bm{p}_{K^+},\bm{p}_{{\cal R}};\bm{p}_{B^0})
\ ,
  \label{eq:vertex2}
\end{eqnarray}
where ``states'' ${\cal R}$ have been introduced
just for conveniently representing $J^P$ of the $D^{*-} D^{*0}$ pair;
${\cal R}$ is not a propagating state.
We consider $J^P=0^+$ and $s$-wave for the $D^{*-} D^{*0}$ pair;
the other $J^P$ does not change the main conclusion
which is essentially determined by the TS.

\begin{figure}[t]
\begin{center}
\includegraphics[width=.5\textwidth]{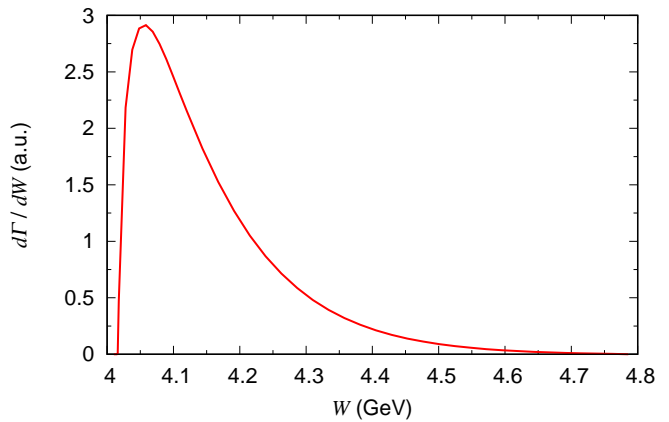}
\end{center}
 \caption{$W$-dependence of the $B^0\to D^{*-} D^{*0} K^+$ differential decay width
from the decay vertex of Eq.~(\ref{eq:vertex2}).
No rescattering is considered.
 }
\label{fig:bddk}
\end{figure}

We introduced the exponential factor in Eq.~(\ref{eq:vertex2}) 
where the parameter $b$ characterizes 
the $W$-dependence of the vertex. 
Although there is no experimental information to fix
the $W$-dependence,
possibly related information is available from other processes such as
the $M_{D^*\bar{D}^*}$ distributions from 
$e^+e^-\to J/\psi D^*\bar{D}^*$~\cite{belle-pakhlov}
and $e^+e^-\to (D^*\bar{D}^*)^\pm\pi^\mp$~\cite{bes3_z4020};
both data show significant enhancements near the 
$D^*\bar{D}^*$ threshold.
We assume that 
the $M_{D^*\bar{D}^*}$ distribution of $B^0\to D^{*-} D^{*0} K^+$
is similar to these data, and that the 
$D^{*-} D^{*0}$ $s$-wave decay vertex of Eq.~(\ref{eq:vertex2})
dominates in the whole $W$ region. 
Then we can fix the parameter $b$ in Eq.~(\ref{eq:vertex2})
as $b=30$ for the cutoff $\Lambda=1$~GeV.
The resulting $M_{D^*\bar{D}^*}$ distribution
is shown in Fig.~\ref{fig:bddk}.
After fixing the $W$ dependence, we can determine
the $B^0\to D^{*-} D^{*0} K^+$ vertex strength using data for the
branching ratio:
${\cal B} (B^0\to D^{*-} D^{*0} K^+)=1.06\pm 0.03\, ({\rm stat.})\pm 0.086\, ({\rm syst.})$\%~\cite{babar_BDDK}.
Since the triangle diagrams in Fig.~\ref{fig:diag} hit
TSs only at $W \sim m_{D^{*-}}+m_{D^{*0}}$,
the $W$-dependence is unimportant for thsee proceeses
in the TS region.
We note that the $X(3872)$ mass determination method~\cite{guo_x3872,sakai_x3872,sakai2_x3872}
analyzes the $W$-dependence only near and in the TS region.
However, the $M_{J/\psi\pi^+\pi^-}$ lineshape
 for $B^0\to (J/\psi\pi^+\pi^-) K^+ \pi^-$ from the Belle~\cite{belle_x3872kpi}
includes data from the whole $W$ region.
Because the $W$-integrated $M_{J/\psi\pi^+\pi^-}$ lineshape 
depends on the $W$-dependence, 
we manage it as above
in order to compare the model with the data.

We evaluate 
the interactions of Eqs.~(\ref{eq:contact}) and (\ref{eq:vertex})
in the CM frame of the two-body subsystem,
and then 
multiply kinematical factors to account for the Lorentz
transformation to the
$J/\psi\rho^0\pi^-$ CM frame; see
Appendix~C of Ref.~\cite{3pi} for details.

\begin{figure*}[t]
\begin{center}
\includegraphics[width=1\textwidth]{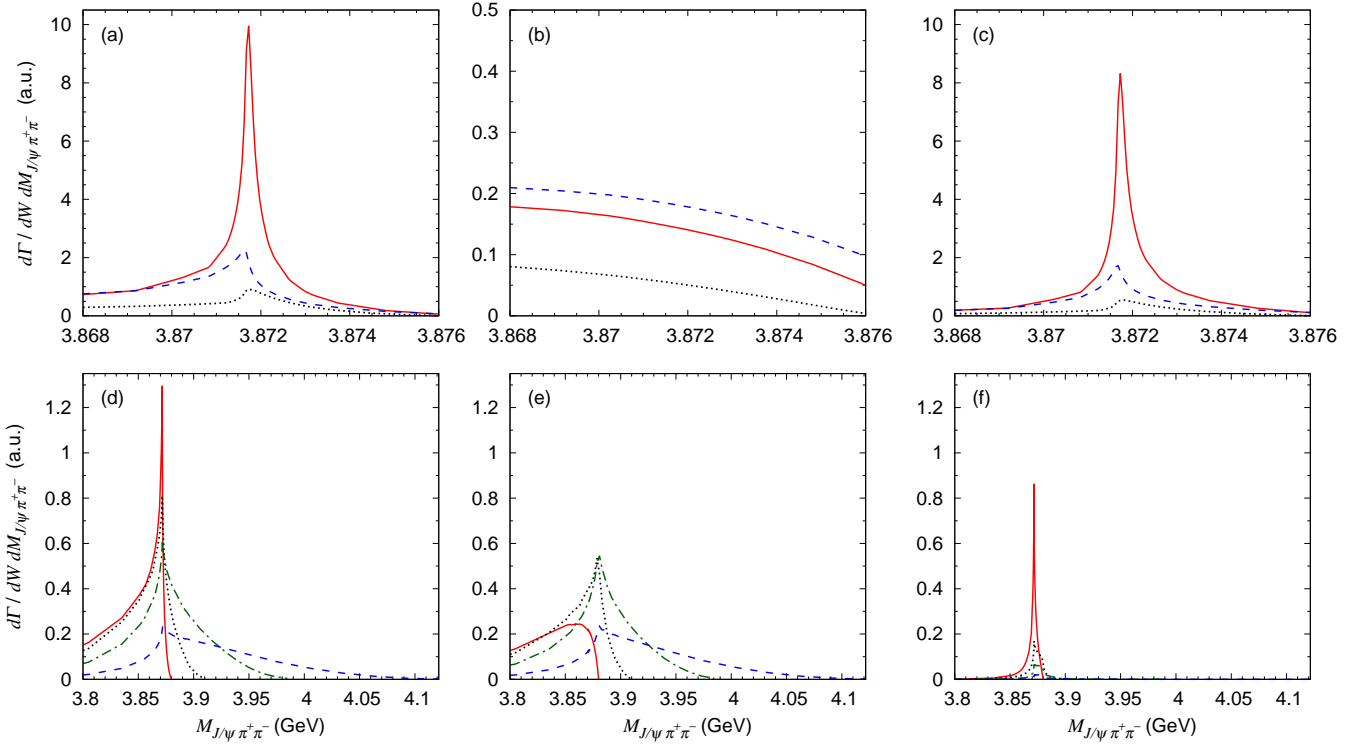}
\end{center}
\caption{
 $J/\psi\pi^+\pi^-$ invariant mass ($M_{J/\psi\pi^+\pi^-}$)
 distributions from the triangle diagrams of Figs.~\ref{fig:diag}(A)
 and \ref{fig:diag}(B) 
for $B^{0}\to J/\psi\pi^+\pi^-K^+\pi^-$.
 $J/\psi$ is paired with $\pi^+\pi^-$ from $\rho^0$ decay 
 to give $M_{J/\psi\pi^+\pi^-}$.
(Upper) The TS region.
The black dotted, red solid, and blue dashed curves correspond to the
 spectra at 
$W-(m_{D^{*0}}+m_{D^{*-}})=-1.0$, 0.0, and 1.2~MeV, 
respectively, where
 $W$ is the invariant mass of the final $J/\psi\pi^+\pi^-\pi^-$
 subsystem and 
$m_{D^{*0}}+m_{D^{*-}}=4017.1$~MeV.
(Lower) Higher $W$ region.
The red solid, black dotted, green dash-dotted, and blue dashed
 curves correspond to the spectra at 
$W-(m_{D^{*0}}+m_{D^{*-}})=2.7$, 32.8, 112.8, and 258.3~MeV, 
 respectively.
In the upper [lower] panels, 
the triangle diagrams of Figs.~\ref{fig:diag}(A) and \ref{fig:diag}(B)
generate the spectra in the panels
 (a) and (b) [(d) and (e)], 
respectively,
and their coherent sum is given in the panel (c) [(f)].
While the vertical scale is arbitrary, the relative scales among the
 curves in all the panels
 are the model-prediction and not arbitrary.
}
\label{fig:spec}
\end{figure*}

The double differential decay width, 
$d\Gamma_{B\to J/\psi\rho^0 K\pi}/dW dM_{J/\psi\rho}$,
is calculated with the decay amplitude of Eq.~(\ref{eq:amp})
in a standard manner as detailed in 
Appendix~B of Ref.~\cite{3pi}.
We then take account of the $\rho^0\to\pi^+\pi^-$ decay.
Thus the final expression for 
the double differential decay width
is given by
\begin{eqnarray}
 { d\Gamma_{B\to J/\psi\pi^+\pi^-K\pi}
  \over dW dM_{J/\psi\pi^+\pi^-}  }
  &=&
  \int  {d M_{\pi\pi}\over 2 \pi}
  \left.
{ d\Gamma_{B\to J/\psi\rho^0 K\pi} \over dW dM_{J/\psi\rho^0}  }\right|_{m_{\rho^0}=M_{\pi\pi}}
\nonumber\\
  &&  
 \times
 {   [M_{\pi\pi} / E_{\rho^0}]^2 \,
 \Gamma_{\rho^0\to\pi^+\pi^-}(M_{\pi\pi})
  \over
|\tilde{W}-E_{\rho^0}+ {i\over 2} \Gamma_{\rho^0}(M_{\pi\pi})|^2
}
\, ,\nonumber\\
\label{eq:decay}
\end{eqnarray}
where
$\tilde{W}\equiv W-E_{J/\psi}-E_{\pi}$ and
the $\rho^0$ nominal mass ($\bar{m}_{\rho^0}=775$~MeV) is used only in
$E_{\rho^0}=\sqrt{\bar{m}^2_{\rho^0}+\bm{p}^2_{\rho^0}}$.
The total and  partial $\rho^0$ decay widths are denoted by
$\Gamma_{\rho^0}$ and $\Gamma_{\rho^0\to\pi^+\pi^-}$, respectively, and
the $M_{\pi\pi}$ dependence is given by 
$\Gamma_{\rho^0}/\bar\Gamma_{\rho^0} = (q/\bar{q})
(\bar{m}_{\rho^0}/M_{\pi\pi})^2
[ f^{10}_{\pi\pi}(q)/f^{10}_{\pi\pi}(\bar{q})]^2$
where $q$ is the pion momentum in the $\rho^0$ rest frame;
the quantities with bar corresponds to the case of 
$M_{\pi\pi}=\bar{m}_{\rho^0}$ and 
$\bar\Gamma_{\rho^0}=150$~MeV;
the form factor $f^{10}_{\pi\pi}(q)$ has been defined in Eq.~(\ref{eq:ff}).

\section{results}

\subsection{$X(3872)$-like TS peak in
$B^0\to (J/\psi\pi^+\pi^-)K^+\pi^-$}

We show in Figs.~\ref{fig:spec}(a,d), \ref{fig:spec}(b,e), and \ref{fig:spec}(c,f)
the $M_{J/\psi\pi^+\pi^-}$ distributions of
the double differential decay width
$d\Gamma_{B^0\to J/\psi\pi^+\pi^-K^+\pi^-} / dW dM_{J/\psi\pi^+\pi^-}$,
defined in Eq.~(\ref{eq:decay}),
from the triangle diagrams~A, B, 
and A+B, respectively.
The spectra in and near the TS region  ($W\sim m_{D^{*0}}+m_{D^{*-}}$)
are shown in Fig.~\ref{fig:spec}(a-c) where
the prominent feature is a very sharp peak
created by the TS from the diagram~A
at $M_{J/\psi\pi^+\pi^-}\sim 3871.7$~MeV, exactly falling on the
precisely measured $X(3872)$ mass:
$3871.69\pm 0.17$~MeV~\cite{pdg}.
We stress that the peak position and the narrow width
due to the TS is virtually parameter-free.
The cutoff dependence over $\Lambda=0.5-2$~GeV
has been confirmed not to significantly change the position and shape of
the TS peak,
and the other arbitrary parameters
can change only the overall normalization.
This stability stems from the facts that:
(i) the TS dominates because
 the tiny $D^*$ width puts the TS very close to the physical region;
(ii) the TS does not depend on dynamical details.
We also find the acute sensitivity of the TS peak to $W$, 
reflecting the fact that the TS region is within a small window of $W$
as in Eq.~(\ref{eq:W-range}).

Meanwhile, the triangle diagram B
gives smooth lineshapes for $W\sim m_{D^{*0}}+m_{D^{*-}}$
as seen in Fig.~\ref{fig:spec}(b).
This sharp contrast between the diagrams~A and B
is from the fact that 
the TS condition is satisfied by the diagram~A only.
Although one might have expected a smaller enhancement arising from 
the diagram B
because of its proximity to the TS condition,
this is not the case. 
The high selectivity of the TS condition shown here is in part due to
the smallness of the $D^*$ width.
The role played by the diagram B in the coherent sum
shown in Fig.~\ref{fig:spec}(c) 
is to remove the smooth background-like contribution from 
the diagram~A.

The $M_{J/\psi\pi^+\pi^-}$ distributions for 
higher $W$ region are given in Fig.~\ref{fig:spec}(d-f).
Figure~\ref{fig:spec}(d) shows that
the remnant of the TS peak from the diagram~A quickly disappears
as $W$ increases, and the threshold cusp stays 
at $M_{J/\psi\pi^+\pi^-}=m_{D^{*0}}+m_{\bar{D}^{0}}$
in the higher $W$ region.
The cusp height for higher $W$ is shorter due to the $W$ dependence
of the $B^0\to D^{*-} D^{*0} K^+$ decay vertex introduced in 
Eq.~(\ref{eq:vertex2}).
The diagram B also generates similar cusps 
at slightly higher energy of 
$M_{J/\psi\pi^+\pi^-}=m_{D^{*-}}+m_{D^{+}}$
as seen in Fig.~\ref{fig:spec}(e).
Thus the coherent sum leaves just a small differece between
contributions from the diagrams~A and B 
as shown in Fig.~\ref{fig:spec}(f).

\begin{figure}[t]
\begin{center}
\includegraphics[width=.5\textwidth]{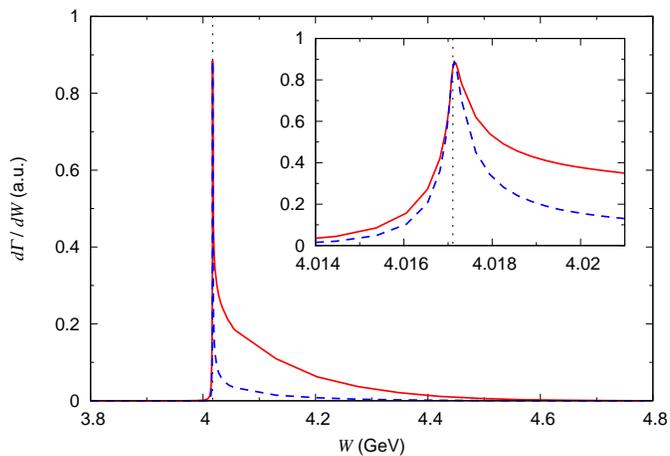}
\end{center}
 \caption{
 $W$ dependence of $B^{0}\to J/\psi\pi^+\pi^-K^+\pi^-$ decay rate
 from the coherent sum of 
 the triangle diagrams of Figs.~\ref{fig:diag}(A) and \ref{fig:diag}(B).
 The red solid curve is obtained by integrating the spectrum in
 Fig.~\ref{fig:spec}(c,f) with respect to $M_{J/\psi\pi^+\pi^-}$
 at each $W$.
 The blue dashed curve is obtained similarly but
 the integral is limited to the range of
 $3.871\le M_{J/\psi\pi^+\pi^-}\le 3.8725$~GeV.
 For a better visibility,
 the blue dashed curve has been scaled by a factor of 1.95.
The vertical line indicates the $D^{*-} D^{*0}$ threshold.
The insert shows the peak region.
 }
\label{fig:w-dep}
\end{figure}

Integrating the spectra in Fig.~\ref{fig:spec}(c,f)
 over $M_{J/\psi\pi^+\pi^-}$ at each $W$ gives 
$d\Gamma_{B^0\to J/\psi\pi^+\pi^-K^+\pi^-} / dW$ which is shown in
Fig.~\ref{fig:w-dep} by the red solid curve.
The spectrum sharply rises and peaks slightly above the $D^{*-}D^{*0}$
threshold, and then falls off.
Because the $W$ dependence of
the spectra in Fig.~\ref{fig:spec}(c) is particularly 
strong around the peak,
we expect an even stronger $W$ dependence of 
$d\Gamma_{B^0\to J/\psi\pi^+\pi^-K^+\pi^-} / dW$
if we limit the integral with respect $M_{J/\psi\pi^+\pi^-}$
to a range around
$M_{J/\psi\pi^+\pi^-}\sim 3871.7$~MeV.
This is indeed the case as shown by
the blue dashed curve in Fig.~\ref{fig:w-dep}.

\begin{figure}[t]
\begin{center}
\includegraphics[width=.5\textwidth]{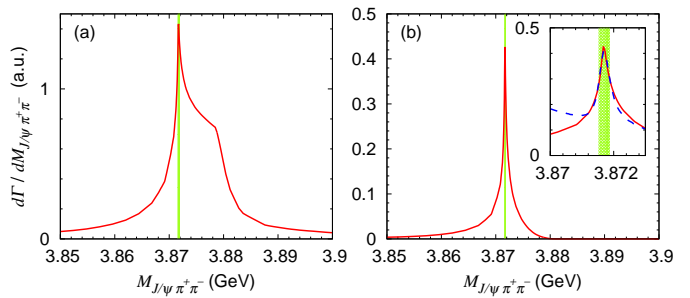}
\end{center}
 \caption{
$M_{J/\psi\pi^+\pi^-}$
 distributions from the triangle diagrams of Figs.~\ref{fig:diag}(A)
 and \ref{fig:diag}(B).
(a) The red solid curve is obtained from the spectra in 
 Fig.~\ref{fig:spec}(c,f) 
 by integrating with respect to $W$
 over the whole $W$-region
as shown in Fig.~\ref{fig:w-dep}.
(b) The red solid curve is obtained similarly, but the integral is
 limited to the range of
$-3~{\rm MeV} \le W - (m_{D^{*-}}+m_{D^{*0}}) \le 4~{\rm MeV}$.
The insert shows the peak region.
 The blue dashed curve is the Breit-Wigner plus a background fitted to the red solid curve.
The green bands indicate the $X(3872)$ mass range from the PDG~\cite{pdg}.
 }
\label{fig:peak}
\end{figure}

The Belle data on
$d\Gamma_{B^0\to J/\psi\pi^+\pi^-K^+\pi^-} / dM_{J/\psi\pi^+\pi^-}$
for  $B^{0}\to J/\psi\pi^+\pi^-K^+\pi^-$~\cite{belle_x3872kpi,belle_x3872kpi2},
showing the peak at $M_{J/\psi\pi^+\pi^-}\sim 3.872$~GeV,
is from the whole kinematically allowed $W$ region. 
To obtain a theoretical counterpart, 
we integrate the spectra in 
Fig.~\ref{fig:spec}(c,f) with respect to $W$.
The resulting spectrum is shown in Fig.~\ref{fig:peak}(a).
We find a sharp peak at $M_{J/\psi\pi^+\pi^-}\sim 3.872$~GeV,
and also a large shoulder near the $D^{*-}D^+$ threshold.
We smeared the spectrum by the experimental resolution, and found that
the lineshape is too broad to be compatible with the data.
This shape depends on the $W$-dependence of the 
$B^0\to D^{*-} D^{*0} K^+$ vertex specified in Eq.~(\ref{eq:vertex2}).
As the higher $W$ region is more suppressed, the shoulder shrinks more.
Although our choice of the $W$-dependence can be different from the
reality to some extent, it seems unlikely that the diagrams~A and B
can explain the Belle data.

\begin{table}[b]
 \caption{\label{tab:BW_param}
 Breit-Wigner mass ($m_{BW}$) and width ($\Gamma_{BW}$).
 The parameters in the second column 
 are determined by fitting the
 $M_{J/\psi\pi^+\pi^-}$ spectrum of Fig.~\ref{fig:peak}(b)
 from the triangle diagrams of Figs.~\ref{fig:diag}(A) and \ref{fig:diag}(B).
 The parameter ranges are from the cutoff dependence ($\Lambda=0.5-2$~GeV).
The PDG values for $X(3872)$ are shown in the third column. 
}
\begin{ruledtabular}
\renewcommand\arraystretch{1.3}
\begin{tabular}{cccc}
& Figs.~\ref{fig:diag}(A)+\ref{fig:diag}(B) & $X(3872)$\ (PDG~\cite{pdg})\\ 
 $m_{BW}$ (MeV)& $3871.68\pm 0.00$& $3871.69\pm 0.17$\\
 $\Gamma_{BW}$ (MeV) & $0.42\pm 0.01$&  $< 1.2$\\
\end{tabular}
\end{ruledtabular}
\end{table}

\begin{figure*}[t]
\begin{center}
\includegraphics[width=1\textwidth]{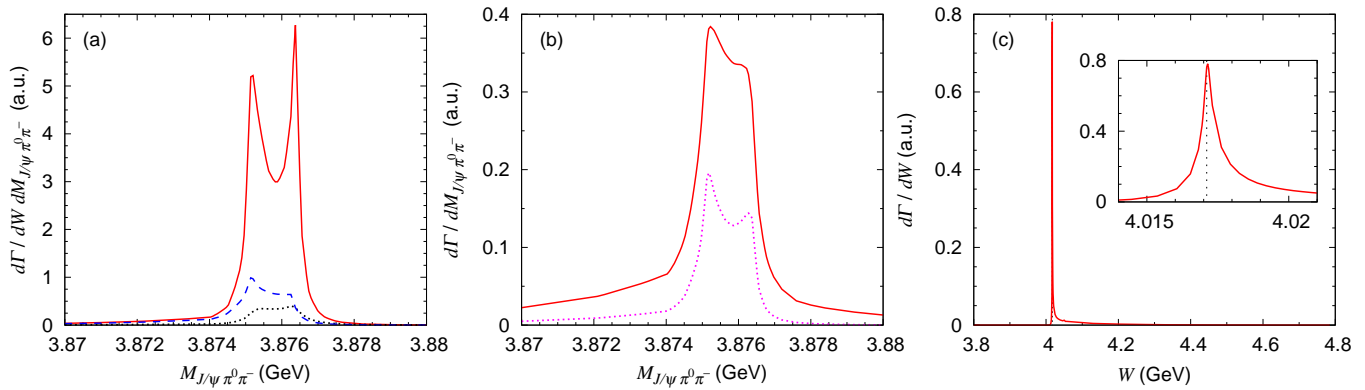}
\end{center}
\caption{
$M_{J/\psi\pi^0\pi^-}$ and $W$ distributions for $B^{0}\to J/\psi\pi^0\pi^-K^+\pi^0$
from the coherently summed triangle diagrams of
 Figs.~\ref{fig:diag}(C) and \ref{fig:diag}(D).
 $J/\psi$ is paired with $\pi^0\pi^-$ from $\rho^-$ decay 
 to give $M_{J/\psi\pi^0\pi^-}$.
(a) The black dotted, red solid, and blue dashed curves correspond to
$M_{J/\psi\pi^0\pi^-}$ distributions at 
$W-(m_{D^{*0}}+m_{D^{*-}})=-1.0$, 0.0, and 1.2~MeV, respectively.
 (b) The red solid curve is obtained from the spectra
 in the panel (a) by integrating over $W$.
The magenta dotted curve is obtained similarly, but the integral range is
 limited to $-3~{\rm MeV} \le W - (m_{D^{*-}}+m_{D^{*0}}) \le 4~{\rm MeV}$.
 (c) The red solid curve is obtained by integrating the spectra in
the panel (a) with respect to $M_{J/\psi\pi^0\pi^-}$ at each $W$.
The vertical line indicates the $D^{*-} D^{*0}$ threshold.
The TS peak region is enlarged in the insert.
Regarding the vertical scale,
the curves in the panels (a), (b), and (c) are comparable to
those in
Figs.~\ref{fig:spec}, \ref{fig:peak}, and \ref{fig:w-dep}, respectively.
}
\label{fig:spec5}
\end{figure*}
We now limit the $W$-integral to near and in the TS region,
$-3~{\rm MeV} \le W - (m_{D^{*-}}+m_{D^{*0}}) \le 4~{\rm MeV}$,
and show the obtained spectrum by the red solid curve in Fig.~\ref{fig:peak}(b).
This time, the narrow peak clearly remains.
To see the peak position and width quantitatively, we simulate the spectrum
with 
the conventional resonance($X$)-excitation mechanism,
$B\to X K\pi$ followed by $X \to J/\psi\rho^0$,
and determine the Breit-Wigner mass and width of $X$.
We also add a coherent background contribution given by
an adjustable quadratic polynomial of $M_{J/\psi\pi^+\pi^-}$.
The result of the fit is shown by the blue dashed curve in
the insert of Fig.~\ref{fig:peak}(b).
The quality of the fit in the tail region
is not very good because: 
(i) the peak shape is rather different from the Breit-Wigner form;
(ii) the background is a quickly decreasing function of 
$M_{J/\psi\pi^+\pi^-}$ near the higher end at 
$\sim 3.88$~GeV due to the limited phase-space.
Still, the obtained Breit-Wigner parameters would be useful to characterize
the peak, and are presented in Table~\ref{tab:BW_param} and compared
with the PDG value for $X(3872)$.
The parameters from the triangle diagrams~A and B are
very stable against changing the cutoff value, and 
in excellent agreement with the precisely measured values for $X(3872)$.
The Breit-Wigner mass value is only a few keV above the $D^{*0}\bar{D}^0$ threshold.
The results indicates that the TS peak 
from the diagram~A could partly fake the $X(3872)$ signal
in $B^{0}\to J/\psi\pi^+\pi^-K^+\pi^-$
around $W\sim m_{D^{*-}}+m_{D^{*0}}$.
This also means that the mechanism 
 could have a possible impact on 
the $X(3872)$ mass determination
method~\cite{guo_x3872,sakai_x3872,sakai2_x3872}.
We will come back to this point later.

\subsection{$X^-(3876)$-like TS peak in
$B^0\to (J/\psi\pi^0\pi^-)K^+\pi^0$}

We now consider a charge analogous process, 
$B^0\to (J/\psi\pi^0\pi^-)K^+\pi^0$,
with the triangle diagrams C and D in Fig.~\ref{fig:diag}.
The $M_{J/\psi\pi^0\pi^-}$ distribution around the TS region 
($W\sim m_{D^{*-}}+m_{D^{*0}}$) is presented in Fig.~\ref{fig:spec5}(a).
The clear twin peaks are created by TSs from 
the triangle diagrams C and D 
at the positions expected in Eqs.~(\ref{eq:ts2}) and (\ref{eq:ts3}), respectively.
Again, the spectra show a sharp $W$ dependence.
Integrating the spectra with respect to $W$, we obtain the red solid
curve in Fig.~\ref{fig:spec5}(b), 
and also the magenta dotted curve that include contributions in and near 
the TS region only. 
The clear peak still remains, and this could appear as an
$X^-(3876)$-like peak in
future data. 
This is an interesting channel to identify a TS in data, because no
resonance(-like) structure similar to this peak has been observed. 
Finally, we integrate
the spectra in Fig.~\ref{fig:spec5}(a) with respect to 
$M_{J/\psi\pi^0\pi^-}$ at each $W$, and present the $W$ distiribution in 
Fig.~\ref{fig:spec5}(c).
The TS also creates a sharp peak here.

\subsection{Possible impact on $X(3872)$ mass determination method using
  $W$-lineshape}

\begin{figure}[t]
\begin{center}
\includegraphics[width=.5\textwidth]{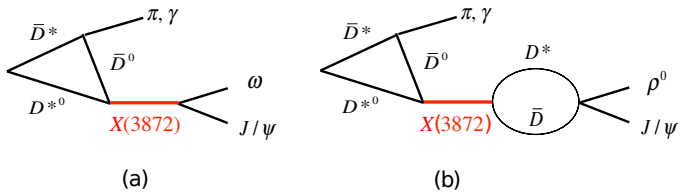}
\end{center}
\caption{
Triangle diagrams utilized in $X(3872)$ mass determination method.
The diagrams include:
(a) isospin-conserving $X(3872)\to J/\psi\omega$ ($\omega\to\pi^+\pi^-\pi^0$);
(b) isospin-violating $X(3872)\to J/\psi\rho^0 (\rho^0\to\pi^+\pi^-)$.
The $D^*\bar{D}$ loop violates the isospin symmetry due to the
 mass difference between the charged and neutral $D^{(*)}$.
}
\label{fig:diag2}
\end{figure}
As stated in the Introduction, 
a method has been proposed recently to determine 
the $X(3872)$ mass by analyzing TS peaks in 
the $X(3872)\pi$ or $X(3872)\gamma$
invariant mass (corresponding $W$) distributions~\cite{guo_x3872,sakai_x3872,sakai2_x3872}. 
The TSs arise from diagrams as shown in Fig.~\ref{fig:diag2} which are
very similar to the diagram~A.
Thus, in the following, 
we speculate a possible impact of
our result, 
Figs.~\ref{fig:spec}(c), 
\ref{fig:peak}(b), 
and Table~\ref{tab:BW_param} in particular,
on this method in a qualitative manner. 
Our result indicates that the TS peak from the diagram~A
would be a perfect fake of $X(3872)$ in the TS region of interest, and
one needs to find a way to extract the $W$-lineshapes from data by separating off the fake from 
$X(3872)$(-like) signal events.

One may wonder if the non-resonant diagram~A would give a negligible
contribution compared with those from the $X(3872)$-pole.
Because $X(3872)$ signal events are reconstructed from its decay
products, 
whether or not an isospin violation, and a significant suppression
 associated with it, occurs in the decay is a key to answer this question.
We first note that the non-resonant diagram~A includes 
the $D^{*0}\bar{D}^0$ pair with 
the maximal mixture of isospin 0 and 1 components. 
Thus both $D^{*0}\bar{D}^0\to J/\psi\rho^0$ as in the diagram~A
and $D^{*0}\bar{D}^0\to J/\psi\omega$ proceed as isospin conserving
processes. 
If the isospin is conserved in a resonant process 
as in Fig.~\ref{fig:diag2}(a),
then the TS peak in the $W$-lineshape
would be almost saturated by this process.
The non-resonant diagram~A, with $\rho^0$ replaced by $\omega$, 
may be negligible.
However, 
if $X(3872)$ singnals are from
isospin-violating decay products such as the process shown in  
Fig.~\ref{fig:diag2}(b), 
the resonant process would be significantly suppressed and 
the diagram~A could be relevant.
The degree of the suppression is uncertain, and could be estimated only
model-dependently.
Although some analysis such as 
Ref.~\cite{prd2012_hanhart} (Table~I therein)
seems to indicate that 
the amplitude magnitude of Fig.~\ref{fig:diag2}(b) is 
$\sim 1/4$ of that of Fig.~\ref{fig:diag2}(a),
this does not necessarily means that the suppression due to the isospin
violation is rather moderate $\sim 1/4$.
It may also be the case that 
the $X(3872)\to J/\psi\omega$ coupling is weak and the suppression is large.

Thus, for measuring the $X(3872)\pi$ and $X(3872)\gamma$ lineshapes,
one may be tempted to utilize a process that involves an
isospin-conserving $X(3872)$ decay, thereby avoiding 
the concern about the non-resonant contributions.
However, the experimentally cleanest signal of $X(3872)$ 
is obtained from the isospin-violating $X(3872)\to J/\psi\pi^+\pi^-$ decay,
and this is the best channel to measure the
$X(3872)\pi$ and $X(3872)\gamma$ lineshapes. 
Therefore, subtracting the background due to the non-resonant TS
process like the diagram~A is a practical issue.

An interesting idea to set a constraint on the non-resonant diagram~A is to
analyze the charge analogous $B^0\to (J/\psi\pi^0\pi^-)K^+\pi^0$ process.
As discussed in the previous section,
the diagrams~C and D creates an $X^-(3876)$-like sharp peak of width of
$\sim 2$~MeV, and 
no similar resonance(-like) peak has been 
experimentally observed in the other processes.
Therefore, if a peak as shown in Fig.~\ref{fig:spec5} is found in 
$B^0\to (J/\psi\pi^0\pi^-)K^+\pi^0$ data, 
this peak is likely to be from TSs of 
the diagrams~C and D.
Thus the data can give an ideal constraint on the magnitude of
the diagrams~C and D and, as a consequence,
the diagrams~A and B are also constrained.
With the well-controlled triangle diagrams~A and B,
we can extract the $X(3872)$-pole amplitude
and the corresponding $W$-lineshape in the TS region
from $B^0\to (J/\psi\pi^+\pi^-)K^+\pi^-$ data.
The $X(3872)$ mass can now be assessed with a good control of
the background.

References~\cite{guo_x3872,sakai_x3872,sakai2_x3872} 
proposed to utilize diagrams in which 
internal particles
$D^{*0}\bar{D}^{*0}D^{0}$ form the triangle.
The charge analogous diagrams including this triangle
do not satisfy the TS condition because
$D^{*0}\to D^+\pi^-$ is forbidden at on-shell.
Therefore, the corresponding non-resonant TS mechanism cannot be studied with 
the charge analogous process.
However, once the relative strength and phase between the resonant and
non-resonant TS mechanisms 
for $B^0\to (J/\psi\pi^+\pi^-)K^+\pi^-$ are understood through the procedure
described above, 
they can be brought to other triangle diagrams studied in 
Refs.~\cite{guo_x3872,sakai_x3872,sakai2_x3872}.
Some necessary corrections can be estimated reliably because they are
associated with kinematical differences.

\section{Summary}
We have demonstrated that the triangle singularity (TS) inherent in the
triangle diagram of Fig.~\ref{fig:diag}(A) creates a sharp peak in
the $J/\psi\pi^+\pi^-$ invariant mass distribution
of $B^0\to J/\psi\pi^+\pi^-K^+\pi^-$.
The Breit-Wigner fit of the peak in and near the TS region
results in
the mass $3871.68\pm 0.00$~MeV and width $0.42\pm 0.01$~MeV
which are in perfect agreement
with those of $X(3872)$,
$3871.69\pm 0.17$~MeV and $< 1.2$~MeV,
from the precise measurements.
The result is virtually independent of the uncertainty of
the model parameters involved.
This is because the TS, which does not depend on
 dynamical details,
determines the peak position and the shape.
However, this TS peak does not explain 
the $X(3872)$ peak observed in 
the Belle data for the same process.
This is because the TS region is rather small in the whole phase-space
and, combined with contributions from the other kinematical region, 
the total peak is significantly broader than 
the $X(3872)$-like TS peak.

We also studied a charge analogous 
$B^0\to (J/\psi\pi^0\pi^-)K^+\pi^0$ process.
We found that the triangle diagrams of 
Figs.~\ref{fig:diag}(C) and \ref{fig:diag}(D)
create an $X^-(3876)$-like TS peak with the width of $\sim 2$~MeV.
We argued that this process is useful for studying
and setting a constraint on the TS
mechanisms because the TS contribution would not overlap with a resonant
one. 

We also argued that the TS peak from 
triangle diagrams like Figs.~\ref{fig:diag}(A)
could be a relevant background 
when extracting TS-enhanced $X(3872)\pi$ and $X(3872)\gamma$ lineshapes from data. 
It has been recently proposed that 
the $X(3872)$ mass can be determined by analyzing the lineshapes in the
TS region.
We suggested a procedure to separate the non-resonant and 
$X(3872)$-pole contributions in the TS region.

\begin{acknowledgments}
The author thanks Y. Kato for useful information on the Belle data,
F.-K. Guo for details on his work, and 
M. Mikhasenko for useful comment on the $W$-integral.
The author is also grateful for Prof. H. Peng's encouragements. 
This work is in part supported by 
National Natural Science Foundation of China (NSFC) under contracts 11625523.
\end{acknowledgments}


\begin{thebibliography}{}

\bibitem{qm}
S. Godfrey and N. Isgur,
Mesons in a relativized quark model with chromodynamics,
Phys. Rev. D {\bf 32}, 189 (1985).

\bibitem{belle_x3872_jpsi-rho}
S.K. Choi et al. (Belle Collaboration),
Observation of a narrow charmoniumlike state in exclusive $B^\pm\to K^\pm \pi^+\pi^- J/\psi$ decays,
Phys. Rev. Lett. {\bf 91}, 262001 (2003).

\bibitem{review_swanson}
E.S. Swanson,
The New heavy mesons: A Status report,
Phys. Rept. {\bf 429}, 243 (2006).

 \bibitem{review_voloshin}
M.B. Voloshin,
	 Charmonium,
Prog. Part. Nucl. Phys. {\bf 61}, 455 (2008).

 \bibitem{review_chen}
 H.-X. Chen, W. Chen, X. Liu, and S.-L. Zhu,
The hidden-charm pentaquark and tetraquark states,
Phys. Rep. {\bf 639}, 1 (2016).

 \bibitem{review_hosaka}
 A. Hosaka, T. Iijima, K. Miyabayashi, Y. Sakai, and S. Yasui,
Exotic hadrons with heavy flavors: $X$, $Y$, $Z$, and related states,
PTEP {\bf 2016}, 062C01 (2016).	 
	 
 \bibitem{review_lebed}
R.F. Lebed, R.E. Mitchell, and E.S. Swanson,
Heavy-Quark QCD Exotica,
Prog. Part. Nucl. Phys. {\bf 93}, 143 (2017).

 \bibitem{review_esposito}
A. Esposito, A. Pilloni, and A.D. Polosa,
 Multiquark Resonances,
Phys. Rept. {\bf 668}, 1 (2017).
	 
 \bibitem{review_ali}
A. Ali, J.S. Lange, and S. Stone, 
Exotics: Heavy Pentaquarks and Tetraquarks,
Prog. Part. Nucl. Phys. {\bf 97}, 123 (2017).

 \bibitem{review_guo}
F.-K. Guo, C. Hanhart, U.-G. Mei{\ss}ner, Q. Wang, Q. Zhao, and B.-S. Zou,
Hadronic molecules,
Rev. Mod. Phys. {\bf 90}, 015004 (2018).

 \bibitem{review_olsen}
S.L. Olsen, T. Skwarnicki, and D. Zieminska,
 Nonstandard heavy mesons and baryons: Experimental evidence,
	 Rev. Mod. Phys. {\bf 90}, 015003 (2018).
	 
 \bibitem{review_raphael}
R.M. Albuquerque, J.M. Dias, K.P. Khemchandani, A. Mart\'inez Torres,
	 F.S. Navarra, M. Nielsen, and C.M. Zanetti,
QCD sum rules approach to the $X$, $Y$, and $Z$ states,
J. Phys. G {\bf 46}, 093002 (2019).	


\bibitem{babar_x3872_jpsi-rho}
B. Aubert et al. (BaBar Collaboration), Study of the $B^-\to J/\psi K^-\pi^+\pi^-$ decay and
	measurement of the $B^-\to X(3872)K^-$ branching fraction,
	Phys. Rev. D {\bf 71}, 071103 (2005).

 \bibitem{cdf_x3872_jpsi-rho}
	D. Acosta et al. (CDF II Collaboration),
	Observation of the narrow state $X(3872)\to J/\psi\pi^+\pi^-$ in
	$\bar{p}p$ collisions at $\sqrt{s}$ = 1.96~TeV,
	Phys. Rev. Lett. {\bf 93}, 072001 (2004).

\bibitem{d0_x3872_jpsi-rho}
	V.M. Abazov et al. (D0 Collaboration),
	Observation and properties of the $X(3872)$ decaying to
	$J/\psi\pi^+\pi^-$ in $p\bar{p}$ collisions at $\sqrt{s}$ = 1.96~TeV,
Phys. Rev. Lett. {\bf 93}, 162002 (2004).

 \bibitem{lhcb_x3872_jpsi-rho}
	 R. Aaij et al. (LHCb Collaboration),
	 Observation of $X(3872)$ production in $pp$ collisions at $\sqrt{s}=7$~TeV,
Eur. Phys. J. C {\bf 72}, 1972 (2012).

\bibitem{bes3_x3872_jpsi-rho}
M. Ablikim et al. (BESIII Collaboration),
Observation of $e^+e^-\to\gamma X(3872)$ at BESIII,
Phys. Rev. Lett. {\bf 112}, 092001 (2014).

\bibitem{babar_x3872_jpsi-omega}
P. del Amo Sanchez et al. (BaBar Collaboration),
Evidence for the decay $X(3872)\to J/\psi\omega$,
Phys. Rev. D {\bf 82}, 011101(R) (2010).

\bibitem{bes3_x3872_jpsi-omega}
M. Ablikim et al. (BESIII Collaboration),
Study of $e^+e^-\to \gamma\omega J/\psi$ and Observation of $X(3872)\to \omega J/\psi$,
Phys. Rev. Lett. {\bf 122}, 232002 (2019).

\bibitem{belle_x3872_jpsi-gamma}
	V. Bhardwaj et al. (Belle Collaboration),
	Observation of $X(3872)\to J/\psi\gamma$ and search for $X(3872)\to\psi'\gamma$ in $B$ decays,
Phys. Rev. Lett. {\bf 107}, 091803 (2011).

\bibitem{belle_x3872_ddstar}
	T. Aushev et al. (Belle Collaboration),
	Study of the $B\to X(3872)(\to D^{*0}\bar{D}^0)K$ decay,
	Phys. Rev. D {\bf 81}, 031103(R) (2010).

\bibitem{swanson_molecule}
	E.S. Swanson,
	Short range structure in the $X(3872)$,
	Phys. Lett. B {\bf 588}, 189 (2004).
	
\bibitem{zhao_molecule}
L. Zhao, L. Ma, and S.-L. Zhu,
Spin-orbit force, recoil corrections, and possible $B\bar{B}^*$ and $D\bar{D}^*$ molecular states,
Phys. Rev. D {\bf 89}, 094026 (2014).

\bibitem{suzuki}
M. Suzuki,
The $X(3872)$ boson: Molecule or charmonium,
Phys. Rev. D {\bf 72}, 114013 (2005).

\bibitem{Kalashnikova}
	Yu.S. Kalashnikova,
	Coupled-channel model for charmonium levels and an option for $X(3872)$,
	Phys. Rev. D {\bf 72}, 034010 (2005).

\bibitem{takizawa}
	M. Takizawa and S. Takeuchi,
	$X(3872)$ as a hybrid state of charmonium and the hadronic molecule,
	PTEP {\bf 2013}, 093D01 (2013).
	
 \bibitem{lqcd_Prelovsek}
S. Prelovsek and L. Leskovec,
Evidence for $X(3872)$ from $DD^*$ Scattering on the Lattice,
Phys. Rev. Lett. {\bf 111}, 192001 (2013).
	
 \bibitem{lqcd_Padmanath}
	M. Padmanath, C.B. Lang, and S. Prelovsek,
	$X(3872)$ and $Y(4140)$ using diquark-antidiquark operators with
	lattice QCD,
	Phys. Rev. D {\bf 92},  034501 (2015).

\bibitem{tetra_maiani}
	L. Maiani, F. Piccinini, A.D. Polosa, and V. Riquer,
	Diquark-antidiquarks with hidden or open charm and the nature of
	$X(3872)$,
	Phys. Rev. D {\bf 71}, 014028 (2005).

\bibitem{tetra_chen}
	W. Chen and S.-L. Zhu,
	The vector and axial-vector charmonium-like states,
	Phys. Rev. D {\bf 83}, 034010 (2011).

 \bibitem{landau}	 
 L.D. Landau,
On analytic properties of vertex parts in quantum field theory,
 Nucl. Phys. {\bf 13}, 181 (1959).

\bibitem{coleman}
S. Coleman and R.E. Norton,
Singularities in the physical region,
Nuovo Cim. {\bf 38}, 438 (1965).

 \bibitem{s-matrix}
R. J. Eden, P. V. Landshoff, D. I. Olive and J. C. Polkinghorne,
The Analytic S-Matrix,
(Cambridge University Press, Cambridge, England, 1966).

\bibitem{TS-Pc2}
M. Bayar, F. Aceti, F.-K. Guo, and E. Oset,
A Discussion on Triangle Singularities in the $\Lambda_b\to J/\psi K^- p$ Reaction,
Phys. Rev. D {\bf 94}, 074039 (2016).

\bibitem{ts1_z3900}
Q. Wang, C. Hanhart, and Q. Zhao,
Decoding the riddle of $Y(4260)$ and $Z_c(3900)$,
Phys. Rev. Lett. {\bf 111}, 132003 (2013).

\bibitem{ts2_z3900}
X.-H. Liu and G. Li,
Exploring the threshold behavior and implications on the nature of $Y(4260)$ and $Z_c(3900)$,
Phys. Rev. D {\bf 88}, 014013 (2013).
	  
\bibitem{szczepaniak}
A.P. Szczepaniak,
Triangle Singularities and $XYZ$ Quarkonium Peaks,
Phys. Lett. B {\bf 747}, 410 (2015).
	
\bibitem{xhliu1}
X.-H. Liu, M. Oka, and Q. Zhao,
Searching for observable effects induced by anomalous triangle singularities,
Phys. Lett. B {\bf 753}, 297 (2016).

\bibitem{xhliu2}
X.-H. Liu and G. Li,
Could the observation of $X(5568)$ be a result of the near threshold rescattering effects?,
Eur. Phys. J. C {\bf 76}, 455 (2016).

\bibitem{xhliu3}
X.-H. Liu,
How to understand the underlying structures of $X(4140)$, $X(4274)$, $X(4500)$ and $X(4700)$,
Phys. Lett. B {\bf 766}, 117 (2017).

\bibitem{ts4_z3900}
A. Pilloni, C. Fernandez-Ramirez, A. Jackura, V. Mathieu, M. Mikhasenko,
	J. Nys, and A.P. Szczepaniak,
Amplitude analysis and the nature of the $Z_c(3900)$,
Phys. Lett. B {\bf 772}, 200 (2017).

\bibitem{ts3_z3900}
Q.-R. Gong, J.-L. Pang, Y.-F. Wang, and H.-Q. Zheng,
The $Z_c(3900)$ peak does not come from the “triangle singularity”,
Eur. Phys. J. C {\bf 78}, 276 (2018).

\bibitem{ts_zc4430}
S.X. Nakamura and K. Tsushima,
$Z_c(4430)$ and $Z_c(4200)$ as triangle singularities,
Phys. Rev. D {\bf 100}, 051502(R) (2019).

\bibitem{ts_z4050}
S.X. Nakamura,
Triangle singularities in $\bar{B}^0\to \chi_{c1}K^-\pi^+$ relevant to $Z_1(4050)$ and $Z_2(4250)$,
Phys. Rev. D {\bf 100}, 011504(R) (2019).

\bibitem{ts_review}
F.-K. Guo, X.-H. Liu, and S. Sakai,
Threshold cusps and triangle singularities in hadronic reactions,
arXiv:1912.07030 [hep-ph].


\bibitem{belle_x3872kpi}
A. Bala et al. (Belle Collaboration),
Observation of $X(3872)$ in $B\to X(3872)K\pi$ decays,
Phys. Rev. D {\bf 91}, 051101(R) (2015).

 \bibitem{belle_x3872kpi2}
 I. Adachi et al. (Belle Collaboration),
Study of $X(3872)$ in $B$ meson decays,
arXiv:0809.1224 [hep-ex].
	 

\bibitem{guo_x3872}
F.-K. Guo, 
Novel Method for Precisely Measuring the $X(3872)$ Mass,
Phys. Rev. Lett. {\bf 122}, 202002 (2019).

\bibitem{sakai_x3872}
S. Sakai, E. Oset, and F.-K. Guo,
Triangle singularity in the $B^-\to K^-\pi^0 X(3872)$ reaction and
	sensitivity to the $X(3872)$ mass,
Phys. Rev. D {\bf 101}, 054030 (2020).

\bibitem{sakai2_x3872}
S. Sakai, H.-J. Jing, and F.-K. Guo, 
Possible precise measurements of the $X(3872)$ mass with the 
$e^+e^-\to \pi^0\gamma X(3872)$ and $p\bar{p}\to \gamma X(3872)$,
arXiv:2008.10829 [hep-ph].

\bibitem{ohio1}
E. Braaten, L.-P. He, and K. Ingles,
Triangle Singularity in the Production of $X(3872)$ and a Photon in $e^+e^−$ Annihilation,
Phys. Rev. D {\bf 100}, 031501(R) (2019).

\bibitem{ohio2}
E. Braaten, L.-P. He, and K. Ingles,
Production of $X(3872)$ Accompanied by a Pion in $B$ Meson Decay,
Phys. Rev. D {\bf 100}, 074028 (2019).

\bibitem{ohio3}
E. Braaten, L.-P. He, and K. Ingles,
Production of $X(3872)$ Accompanied by a Soft Pion at Hadron Colliders,
Phys. Rev. D {\bf 100}, 094006 (2019).

 \bibitem{pdg}
M. Tanabashi et al. (Particle Data Group),
Review of Particle Physics,
Phys. Rev. D {\bf 98}, 030001 (2018).

\bibitem{rosner_dstar}
J.L. Rosner,
Hadronic and radiative $D^*$ widths,
Phys. Rev. D {\bf 88}, 034034 (2013).

\bibitem{belle-pakhlov}
P. Pakhlov et al. (Belle Collaboration),
Production of New Charmoniumlike States in $e^+e^-\to J/\psi D^{(*)} \bar{D}^{(*)}$
at $\sqrt{s} \approx 10.6$~GeV,
Phys. Rev. Lett. {\bf 100}, 202001 (2008).

 \bibitem{bes3_z4020}
 M. Ablikim et al. (BESIII Collaboration),
	 Observation of a charged charmoniumlike structure in
$e^+e^-\to (D^*\bar{D}^*)^\pm\pi^\mp$ at $\sqrt{s}=4.26$~GeV,
Phys. Rev. Lett. {\bf 112}, 132001 (2014).
	 
\bibitem{babar_BDDK}
P. del Amo Sanchez (BaBar Collaboration),
Measurement of the $B\to \bar{D}^{(*)}D^{(*)}K$ branching fractions,
Phys. Rev. D {\bf 83}, 032004 (2011).
	 
 \bibitem{3pi}
H. Kamano, S.X. Nakamura, T.-S.H. Lee, and T. Sato,
Unitary coupled-channels model for three-mesons decays of heavy mesons,
Phys. Rev. D {\bf 84}, 114019 (2011).

\bibitem{prd2012_hanhart}	
C. Hanhart, Yu.S. Kalashnikova, A.E. Kudryavtsev, and A.V. Nefediev,
Remarks on the quantum numbers of $X(3872)$ from the invariant mass
distributions of the $\rho J/\psi$ and $\omega J/\psi$ final states,
Phys. Rev. D {\bf 85}, 011501(R) (2012).
	
\end{thebibliography}


\end{document}